\documentclass{cta-author}
{}
{}
{}
\usepackage{etoolbox}
\makeatletter
\def\ps@pprintTitle{%
   \let\@oddhead\@empty
   \let\@evenhead\@empty
   \let\@oddfoot\@empty
   \let\@evenfoot\@oddfoot
}
\makeatother
\usepackage{placeins}
 \usepackage{url}
 \usepackage{comment}
\usepackage{listings}
\usepackage{setspace}
\usepackage{xcolor,pict2e}
\definecolor{eclipseStrings}{RGB}{42,0.0,255}
\definecolor{eclipseKeywords}{RGB}{127,0,85}
\colorlet{numb}{magenta!60!black}

\lstdefinelanguage{json}{
    basicstyle=\normalfont\ttfamily,
    commentstyle=\color{eclipseStrings}, 
    stringstyle=\color{eclipseKeywords}, 
    numbers=left,
    numberstyle=\scriptsize,
    stepnumber=1,
    numbersep=8pt,
    showstringspaces=false,
    breaklines=true,
    frame=lines,
    backgroundcolor=\color{white}, 
    string=[s]{"}{"},
    comment=[l]{:\ "},
    morecomment=[l]{:"},
    literate=
        *{0}{{{\color{numb}0}}}{1}
         {1}{{{\color{numb}1}}}{1}
         {2}{{{\color{numb}2}}}{1}
         {3}{{{\color{numb}3}}}{1}
         {4}{{{\color{numb}4}}}{1}
         {5}{{{\color{numb}5}}}{1}
         {6}{{{\color{numb}6}}}{1}
         {7}{{{\color{numb}7}}}{1}
         {8}{{{\color{numb}8}}}{1}
         {9}{{{\color{numb}9}}}{1}
}

\begin{document}

\title{IoT Monitoring with Blockchain: Generating Smart Contracts from Service Level Agreements}

\author{\au{Adam Booth$^{1\corr}$}, \au{Awatif Alqahtani$^{2\corr}$}, \au{Ellis Solaiman$^{1\corr}$}}
\address{\add{1}{Computing School, Newcastle University, Newcastle, UK}
\email{a.booth2@newcastle.ac.uk, ellis.solaiman@ncl.ac.uk}\\
\add{2}{Computer Science and Engineering, College of Applied Studies and Community Service, King Saud University, Riyadh, SA}
\email{aqahtani1@ksu.edu.sa}\\}

\begin{abstract}
A Service Level Agreement (SLA) is a commitment between a client and provider that assures the quality of service (QoS) a client can expect to receive when purchasing a service. However, evidence of SLA violations in Internet of Things (IoT) service monitoring data can be manipulated by the provider or consumer, resulting in an issue of trust between contracted parties. The following research aims to explore the use of blockchain technology in monitoring IoT systems using smart contracts so that SLA violations captured are irrefutable amongst service providers and clients. The research focuses on the development of a Java library that is capable of generating a smart contract from a given SLA. A smart contract generated by this library is validated through a mock scenario presented in the form of a Remote Patient Monitoring IoT system. In this scenario, the findings demonstrate a 100 percent success rate in capturing all emulated violations.
\footnote
 This paper is a postprint of a paper submitted to and accepted for publication in 
[journal] and is subject to Institution of Engineering and Technology Copyright. The copy of record is 
available at the IET Digital Library
\end{abstract}

\maketitle

\vspace{2cm}
Note:\\  
This paper is a postprint of a paper submitted to and accepted for publication in 
[Managing Internet of Things Applications Across Edge and Cloud Datacenters. IET] and is subject to Institution of Engineering and Technology Copyright. The copy of record is 
available at the IET Digital Library

\vspace{1cm}
\onecolumn

\section{Introduction}
In recent years there has been an uprise in Internet-of-Things (IoT) systems that are now being employed to manage services such as smart energy, smart health care and smart living. Part of this uprise can be attributed to the ease and cost effectiveness of acquiring systems and services that can be used to develop IoT systems from cloud platform providers \cite{1}.\\

When acquiring a service from a third party, a service level agreement (SLA) is agreed upon between the service provider and consumer which details the quality of service (QoS) one can expect to receive. One example could be a cloud-based database service with a QoS rule stating that the service provides a minimum of 1,000 read queries a second. While a service provider will strive to achieve the QoS detailed in the SLA, however, at times the service provider may fail to meet the demands of the SLA which results in an SLA violation. An SLA violation often results in the service provider being penalised in the form of a refund in service credit.\\

Detecting SLA violations can be achieved by monitoring the performance of a service against the QoS requirements of the SLA. It is argued that a single party should not be responsible for this monitoring \cite{25}, yet challenges and trust issues arise when multiple parties are monitoring a system due to the possibility of foul play. This issue of trust arises from the possibility of a party being able to tamper with the monitoring data they have captured to give a false account of the services performance. For example, if the service provider is monitoring the service and the captured data shows they have committed a violation, they could alter these logs and deny the occurrence. Alternatively, the service consumer may tamper with the monitoring data they have captured to falsely accuse the service provider of committing an SLA violation. With parties being able to tamper with the monitoring data they have captured, it is therefore difficult to achieve trust between parties and to determine what has actually occurred.
One solution to solve this issue is to recruit a trusted third party (TTP) to monitor the service on behalf of both the service provider and consumer. However, as mentioned by Aniello et al., this solution results in added cost to the system along with a potential single point of failure \cite{21}. Therefore, rather than using a TTP a better solution may be to use blockchain technology and smart contracts to monitor services to detect SLA violations.\\

Blockchain technology provides a distributed, immutable data store such that once data has been recorded on to the blockchain it is extremely difficult to alter it. Along with this data immutability, blockchain removes the possibility of a single point of failure due to the distributed nature of the technology. This aspect of distribution and immutability could remove issues of trust among parties as they would not be able to tamper with the data recorded and can take part in maintaining the blockchain. However, blockchain alone is not enough as logic is still required to evaluate the performance of a service to determine if an SLA violation has occurred.
Some blockchain platforms such as Ethereum and Hyperledger Fabric provide a runtime environment for smart contracts, allowing for applications to be executed on top of the blockchain. Therefore, a smart contract could be developed for a specific SLA and deployed to a blockchain platform to monitor a service autonomously. This would result in a solution to monitoring services for SLA violations in a trustworthy and irrefutable fashion.\\

While a person could manually develop a smart contract to monitor a service for a given SLA, a better alternative may be to develop software that can interpret an SLA and automatically generate a monitoring smart contract. Research such as \cite{10} \cite{25} have provided formal tools and languages that can define SLAs in a non-ambiguous format allowing for software to interpret these SLAs and extract various requirements from them. Alqahtani et al. have developed an end-to-end IoT SLA generation tool such that a person can define an SLA for an entire IoT system \cite{10}. SLAs produced by this tool could be interpreted by a piece of software to automatically generate a smart contract capable of monitoring the entire IoT system. The aim of this research is to develop software that can achieve this goal of generating a smart contract for monitoring an entire IoT system from an SLA produced by the end-to-end IoT SLA generation tool \cite{10}. \\

The software produced in this  research is a Java programming library that can interpret an SLA and produce a smart contract for monitoring an entire IoT system. The smart contracts generated by this library target the Hyperledger Fabric blockchain platform as this platform provides the ability to deploy a consortium blockchain network which removes unwanted side effects such as publicly available data and transaction costs that is not as easily achieved when using a public blockchain. We aim to reflect the effectiveness of deploying a smart contract on blockchain in automating SLA monitoring in a more trustworthy fashion. Thus,  we  develop a Java library that can extend our work in by converting the generated  SLAs  to a smart contract that is deployable to a blockchain platform and is capable of monitoring an entire IoT system for SLA violations. \\

\textbf{contribution}
\begin{itemize}
\item Develop a Java library such that it can generate smart contracts capable of monitoring an IoT system for SLA violations.
\item Evaluate the correctness of the smart contracts generated by comparing the smart contracts against the original service level agreements.
\item Deploy the generated smart contract onto a Hyperledger Fabric blockchain network.
\item  Evaluate the performance of the smart contract to reflect its effectiveness in capturing SLA violation in autonomously.
\end{itemize}
\section{Background}
\subsection{Internet of Things}
IoT systems are growing in popularity with the expected number of IoT devices to exceed 50 billion by the year 2025 \cite{r2}. As research and developments increase in IoT, more services and devices are being deployed that are used in everyday life such as transport, healthcare and smart living. As reliance and responsibility is placed on these systems developed, it is important to ensure that these IoT systems function correctly and do not fail or degrade. \\

An IoT system typically consists of multiple layers, with each layer providing some sort of service and functionality to the system. These layers often take advantage of services that are either provided by third-party vendors or developed by the organisation that is creating the system. For example, if an organisation were to develop a smart patient monitoring system, the organisation may develop their own devices that perform the physical monitoring of the patient while acquiring a cloud-based data analytics service from a third party to analyse the data collected. 
\FloatBarrier
\begin{figure}[h!]
	\center{\includegraphics[height=5cm]{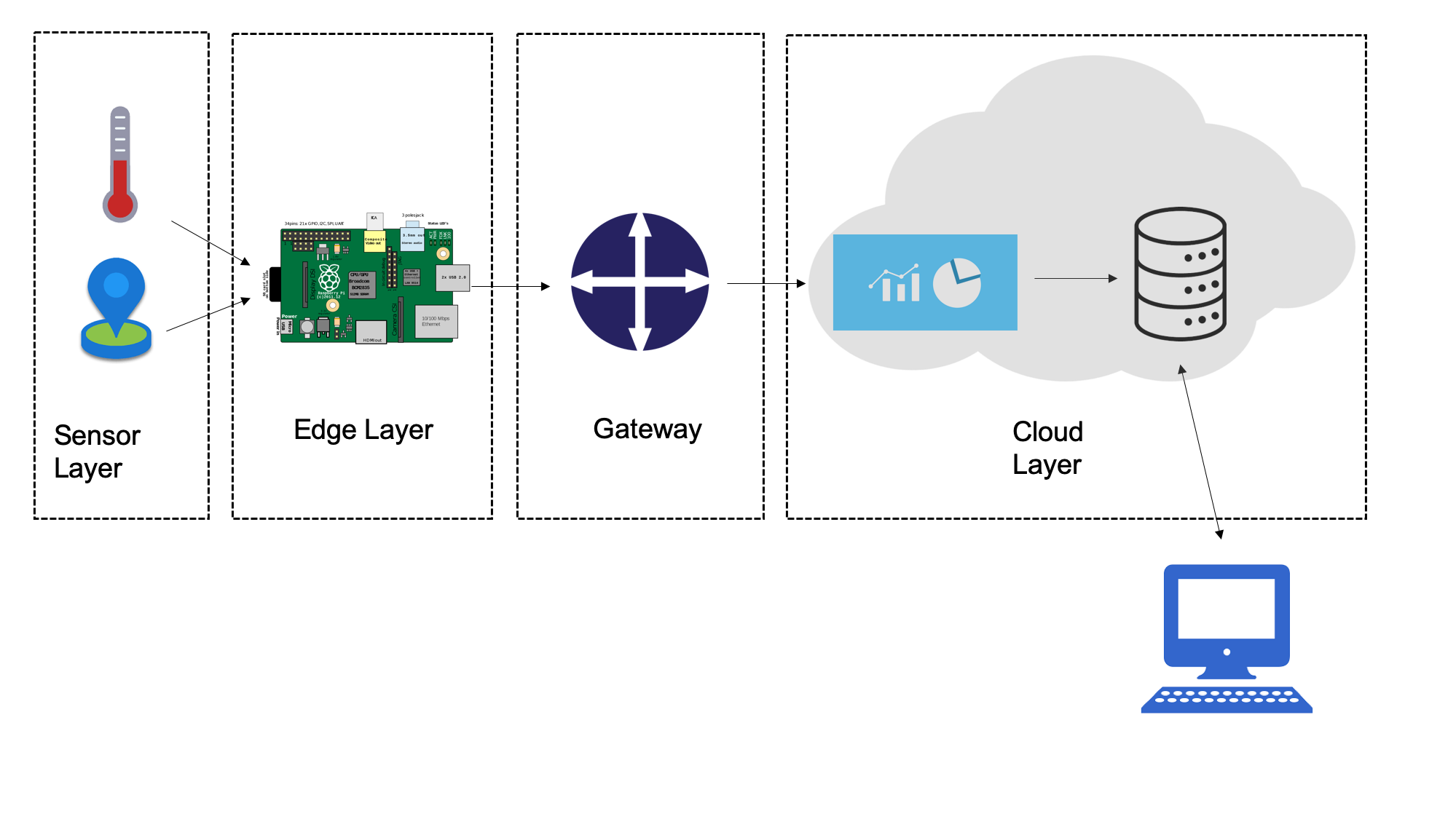}}
	\caption{\label{fig:iotsys} Simple IoT System}
\end{figure} 
\FloatBarrier
Figure \ref{fig:iotsys} shows a simple IoT system  illustrating the various layers. The first layer is the sensor layer, which in this case is responsible for sensing the environment for the  temperature and location. The second layer is the edge layer, which is responsible for collecting data from these sensors and performing some operations on the data such as compression. This compressed data is then provided to a data analytics service (cloud layer) via networking services (gateway layer). Once the data has been received and analysed by the data analytics service, the results are stored in a database on the cloud. A user is then able to query this database for statistics and data analytics. \\

Due to the structure of a typical IoT system, each layer is often relying on some other layer in the system in order to achieve the overall goal. This nature of relying on other layers and services within an IoT system in turn requires monitoring for each service used to ensure that all services are running optimally. This is required as if a layer or service fails or severely degrades in performance, it may have an impact on the overall system. For example, if all gateway services that are responsible for routing data to a data anlytics service fails, then this collected data will not be received and the data may be lost. There has been much research into monitoring cloud-based services and IoT systems such as \cite{r3,r4,r5} with many tools being developed to monitor specific aspects of a system.
\subsection{Service Level Agreements}
A service level agreement (SLA) is an agreement made between a service provider and service consumer that expresses the various Quality of Service (QoS) properties of the service to be provided. For example, a cloud-based database service may include rules in the SLA such as "the availability will always be above 99.999 per cent" and " the database is able to achieve a minimum of 1000 database queries per second". The SLA therefore allows a service consumer to determine if the service is capable of achieving the various performance requirements that their IoT system requires.\\

While service providers who have agreed to an SLA will strive to meet at least the minimum requirements of the agreement, sometimes the provider may fail to achieve them. This failure is known as an SLA violation. When an SLA violation occurs, often the service provider will be penalised for not meeting the service requirements. This penalty is often defined with the SLA, outlining the process of filing a report of a violation and what the penalty will be based on the situation. A service provider will often provide some form of refund as a penalty when an SLA violation has occurred.
\subsection{Blockchain}
Blockchain is effectively an indelible, distributed ledger such that once a piece of data or a transaction has been written to the ledger it cannot be edited or removed. The first widely used blockchain platform was the cryptocurrency Bitcoin, developed under the pseudonym of Satoshi Nakamoto \cite{12}. 
Two other classifications of blockchain exists which are private and consortium blockchains. A private blockchain is typically owned and utilised by a single organisation and only members of that organisation may read/write to the blockchain. Private blockchains are the most restrictive in terms of access and is often internal to a single organisation \cite{14}.\\

A consortium blockchain is often owned and utilised by two or more organisations and members of the consortium take part in administrating the blockchain along with taking part in validation and consensus. Members are able to access the ledger whilst preventing the general public from being able to access the blockchain through various security measures and access policies. Therefore, consortium Blockchains are particularly useful when multiple organisations wish to have a private blockchain that all members of a consortium can access \cite{14}.
The most important aspects of a blockchain network is its validation process and consensus algorithm. Validation is the process of validating submitted transactions to ensure that the transaction is non-malicious and prevent issues such as double spending \cite{15}. Valid transactions are grouped together to form a block which is then appended to the blockchain via a consensus protocol.\\

\begin{figure}[h!]
	\center{\includegraphics[width=\textwidth]{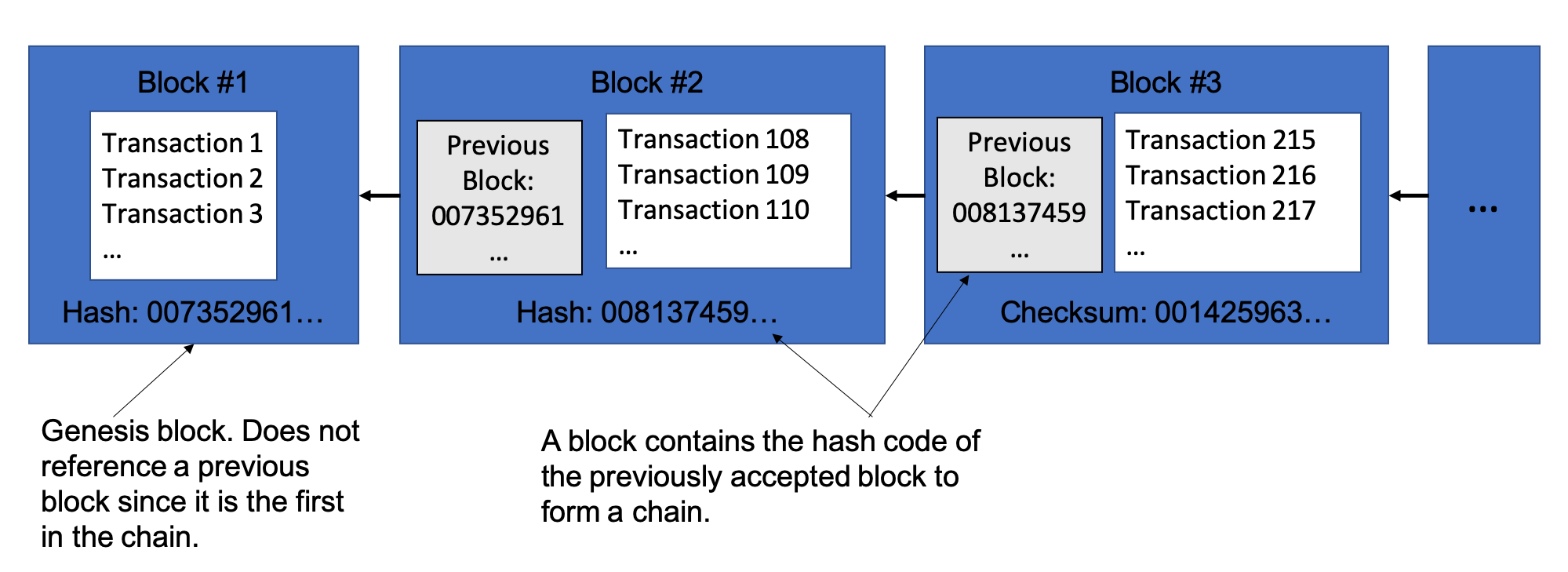}}
	\caption{\label{fig:blockchain} blockchain structure}
\end{figure}

While different blockchain platforms may utilise different consensus algorithms and may restrict read/write access, the formation of a blockchain is typically the same. A block is generated from a group of transactions that is validated by members of the blockchain. Once a block is validated it is then added to the blockchain by following the consensus protocol in use.
A block is appended to a blockchain by including the hash value of the previous block, which is derived by using some form of hashing algorithm such as SHA256. A hashing algorithm computes a fixed length value based on a block of data such that if a single bit were to be changed in the block of data, a different hash value would be generated. By including the hash value of the previously generated block in the newly generated block, this forms a chain linking each block to its predecessor\cite{12}.This process of including the hash value of the previous block in turn provides immutability to the blockchain. 
Figure \ref{fig:blockchain}  illustrates this linkage between blocks in a blockchain.
\FloatBarrier
\FloatBarrier
\subsection{Smart Contracts}
Smart contracts are computer programs that are implementations of formal contracts agreed to by two or more parties\cite{17}.A smart contract is responsible for detecting events and deviations of a contract and automatically carrying out actions based on those events captured \cite{18}. For example, if a person who has agreed to a contract and a clause of that contract states the person must provide some service by a certain date and fails to do so, the smart contract would detect this violation of the contract and issue some form of fine defined in the contract autonomously.\\

While smart contracts may provide an efficient and convenient way of mediating a contract, however, some issues exist in gaining trust of the smart contract between parties. Firstly, if a smart contract is running on a single machine, this results in a single point of failure such that if the computer fails, the smart contract will not capture any events and will therefore not carry out the various required actions, thus failing to meet the contract requirements. Another issue is trusting that the smart contract will not be tampered with. For example, if a party member owns the machine that is running the smart contract or has some other form of access to it; they may be able to modify the code of the smart contract in order to benefit in some way.\\

The above issues of smart contracts, however, can be avoided using blockchain technology. Some blockchain platforms such as Ethereum and Hyperledger Fabric provide the ability to deploy a smart contract to the blockchain, such that all members taking part in the validation and consensus protocol execute the contract and record the results. Since the smart contract is distributed on the blockchain, there is no single point of failure and it can be trusted that the smart contract will be executed.
In addition, due to the immutable nature of blockchain technology once a smart contract has been deployed to the blockchain the code of the contract cannot be tampered with. Therefore, parties that have agreed to the smart contract can trust that the smart contract will remain unmodified. Additionally, if the smart contract writes records such as logs, records and transactions to the distributed ledger, one can trust that the data produced by the smart contract is trustworthy and has not been tampered with.
\subsection{Hyperledger Fabric}
Hyperledger Fabric is an open source implementation of a blockchain framework which includes a modular architecture such that blockchain developers/admins can use various different consensus protocols and membership options to deploy a blockchain network \cite{19}. Hyperledger Fabric also includes a smart contract engine that is capable of executing smart contracts written in Java, JavaScript and Go. Hyperledger Fabric allows the development and deployment of consortium blockchains by providing a set of tools that allows an administrator to generate various certificates and configuration files that are then used to deploy a blockchain network.Due to Fabrics plug-and-play architecture, it allows an administrator to select various database technologies to act as the data store such as ‘LevelDB’ and ‘CouchDB’ along with allowing the administer to specify membership policies and consensus protocols to use.\\

In Hyperledger Fabric, consensus is achieved in three stages which are the endorsement, ordering and validation stages. Endorsement is achieved such that when a transaction is submitted, X out Y nodes must endorse the transaction in order for it to be accepted. Once the transaction has been endorsed, the ordering phase begins and accepts endorsed transactions and agrees to the order of transactions that should be committed. Lastly, once a block of transactions has been ordered, validation is carried out on the ordered block and validates the correctness of it. If the block is validated successfully, it is accepted and appended to the blockchain \cite{20}. Hyperledger Fabric allows a blockchain administrator to specify various services to use for each of the three phases, allowing for a highly configurable consensus protocol. Hyperledger Fabric is therefore a highly configurable blockchain platform suitable for consortium use. It provides a high level of security and control over various access policies, allows administrators to specify which consensus pro- tocols to use and allows for the deployment and execution of smart contracts written in many programming languages.
\section{Related Work}
In their research Aniello et al. propose a system called SLAVE (Service Level Agreement VErified) which recruits public blockchain technology to embed logs of a system that can be used to prove an SLA violation \cite{21}. Aniello et al. argue that a trusted-third party (TTP) has drawbacks including performance overheads, single point of failure and bottle neck issues and the use of blockchain technology could replace a TTP. While SLAVE is proposed to utilise public blockchain technology, however, the price of running the system in terms of transaction costs is not covered and is a suggested further work worth investigating. Aniello et al. also note that while blockchain may remove the possibility of false claims and accusations, a bottle neck issue still remains in their proposed system.\\

Alzubaidi et al. suggest a blockchain-based SLA management solution in the context of IoT monitoring and state that a single party should not be solely responsible for the control of the SLA and monitoring for violations \cite{25}. Alzubaidi et al. argue that in a complex IoT system, it is difficult to determine who is responsible for a failure in the system due to multiple parties having influence over the functionalities. Therefore, it may not be possible for a single member to achieve a complete view of the system and possess all information. It is proposed that using a blockchain solution for SLA monitoring would remove issues from the current practices by providing automatic conflict resolution, remove of trust upon a single authority and deliver complete awareness of the entire IoT system to all parties involved. Alzubaidi et al. consider consortium blockchains, suggesting that Hyperledger Fabric may be the best blockchain platform as opposed to public blockchains such as Ethereum due to not incurring transaction costs, provides a faster performance and scalable platform.\\

Neidhardt et al. propose an SLA solution that utilises smart contracts and the Ethereum blockchain platform for billing and detecting availability SLA violations \cite{23}. The system proposed utilises a smart contract developed to detect when an acquired service is unavailable which results in a customer having coins deposited to their Ethereum wallet. At the end of each billing cycle, the coins credited to the customer is counted and if the number of coins exceeds a certain value, an availability SLA violation is determined. In this research, Neidhardt et al. propose that the user can pay for the services they have used using Ether, Ethereums currency. While only a prototype, it is argued that this research shows promise to achieve improvement in both the billing process of cloud-based services and checking for SLA violations. However, Neidhardt et al. also state that due to the volatility of the price of Ether, this may make it an impractical solution for billing and service prices.
In their research, Scheid et al. propose that the compensation process of an SLA violation is complex due to the level of manual effort required and that a better solution may be to automate the compensation and payment process using blockchain technology and smart contracts\cite{24}. Scheid et al. argue that there are potential issues for both the service consumer and service provider as the consumer could refuse to pay for a service already used and a provider could refuse to compensate when a violation occurs. To overcome this issue, Scheid et al. develop a smart contract which is deployed to a blockchain platform (Ganache) to monitor a sample application for response time. If the response is too slow, a violation is captured and if the violation is severe enough, it automatically compensates the service consumer.\\

Uriarte et al. in their research explore the possibility of converting dynamic service level agreements into smart contracts \cite{25} In their previous work, a formal SLA definition language called SLAC was developed to define SLA's and the goal of the research in \cite{25}  was to further this research by developing an SLA to smart contract conversion service which the resulting smart contract could be deployed to monitor a service being provided. The service and prototype shown in their research converts an SLA defined using SLAC into a smart contract written in Solidity, targeting the Ethereum blockchain network. While the authors state the prototype is still in development a description of the system is provided. Uriarte et al. argue that their system could disrupt the cloud market by allowing for the notion of an open and distributed cloud, offering dynamic services to customers resulting in reduced cost and increased flexibility.
\section{Design and Implementation}
This section focuses on the design and implementation of the smart contract generation library and aims to provide the reader a full understanding as to how the library generates a smart contract. The data model of the library created in this project is closely modelled on the JSON SLA document produced by the SLA generation tool used. Therefore, a brief description of the SLA schema is first provided and will allow the reader to more easily understand the inner workings of the library created.
\subsection{Service Level Agreement}
\begin{figure*}[!htbp]
      \centering
\includegraphics[width=1\linewidth ]{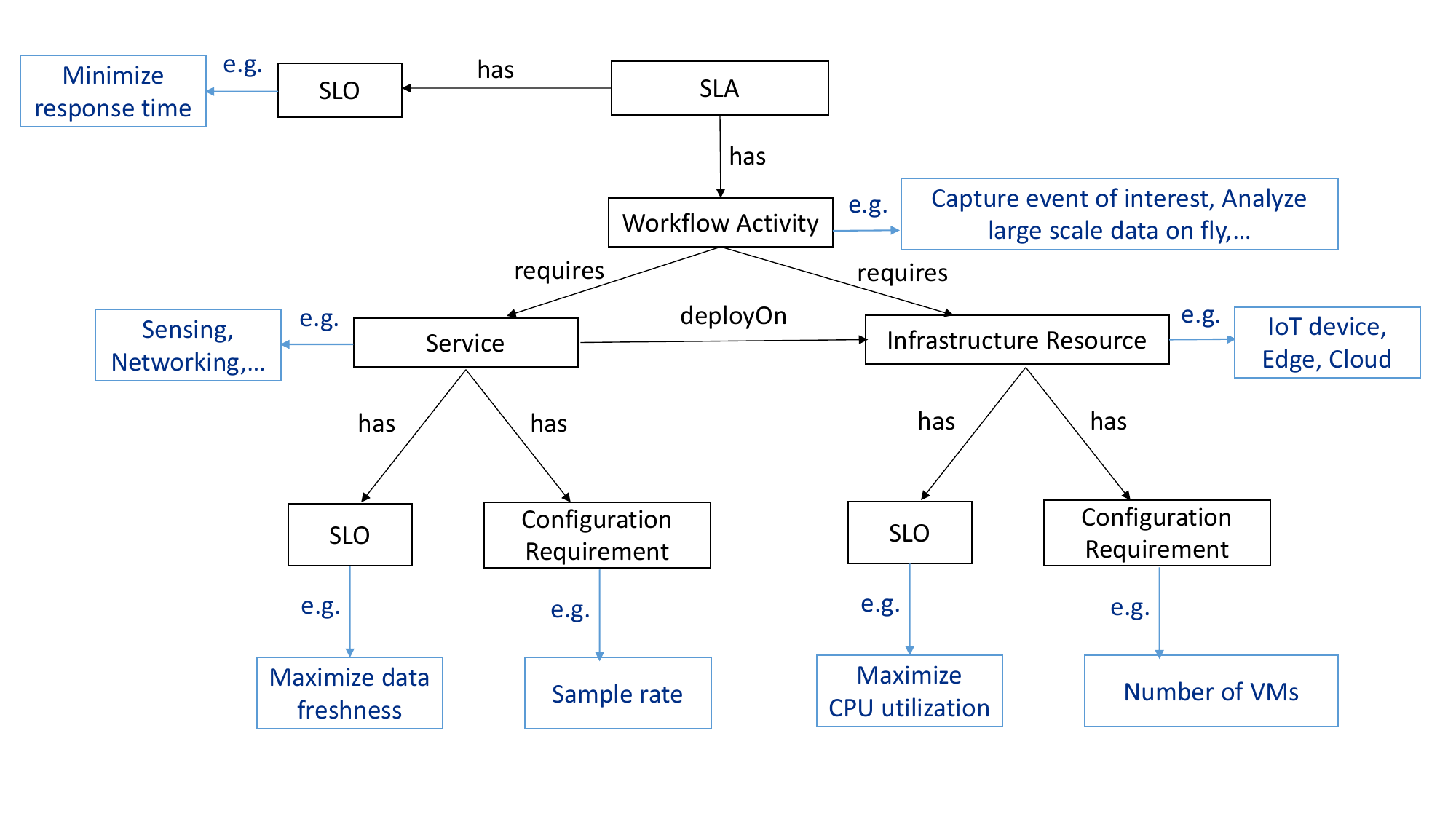}
      \caption{Conceptual model with examples to illustrate the relationships among the key concepts of an SLA for the IoT }
      \label{fig:conceptual11}
   \end{figure*}

Alqahtani et al. have created a tool that allows a user to easily define an end-to-end service level agreement of an IoT system \cite{10}. This is achieved by allowing a user to specify the QoS metrics expected of the application along with the expected QoS of all other layers within the IoT system. There is a number of workflow activities that requires services such as stream processing service as well as an infrastructure resource to host that service. There is a number of service level Objectives (SLOs) associated with the required service and ifrastructure resources. Along with these SLO's, there is other configuration  requirements of the infrastructure resources and services for the involved workflow activities. Figure \ref{fig:conceptual11}  shows the main concepts that has been considered within the SLA with some examples for clarification.

\FloatBarrier
\FloatBarrier

Figure \ref{fig:conceptual11}  associates the concepts presented in the conceptual model with examples to illustrate the relation between infrastructure resources, services, configuration requirement and SLO concepts. For example, "capture event of interest" is a possible workflow activity, in RHMS, and it requires a sensing service. The sensing service has SLO constraints such as the required level of data timeliness. The sensing service will be deployed/hosted on an IoT device. Therefore, it is important to consider the requirements of the IoT device, such as its type (e.g., sensor or RFID), the mobility of the device (e.g., fixed or mobile), the communication mechanism (e.g., pushing data or pulling data) and the battery life. The same conditions will apply for the "filter a captured event of interest" activity, which will be performed at the Edge of a network to filter data and utilize network bandwidth by neglecting uninteresting data; this task uses certain devices, such as a mobile phone or raspberry pi. Each of these devices has specific computational capabilities, such as a given CPU speed and memory size. Furthermore, to perform the "real-time data analysis" activity, a stream processing service can be used with certain requirement constraints, such as low latency and certain configuration requirements, including the specification of the window type as a time-based window or event-based window. The stream processing service can be deployed on a Cloud, so certain requirements related to a Cloud resource can be specified, such as the number of VMs and the acceptable percentage of CPU utilisation. Therefore, in next section, we identify the related vocabulary terms that can be used for specifying QoS and configuration parameters. \\

\subsection{From SLA to Smart Contract Java Library} \label{sec:library}

Generating SLA in a machine-readable format simplifies the process of translating the SLA to Smart contracts. 
 We aimed to translate the generated SLA to a smart contract and explore the use of Blockchain technology to monitor IoT applications to record SLA violations.\\
 
\begin{figure*}[!htbp]
      \centering
\includegraphics[width=1\linewidth]{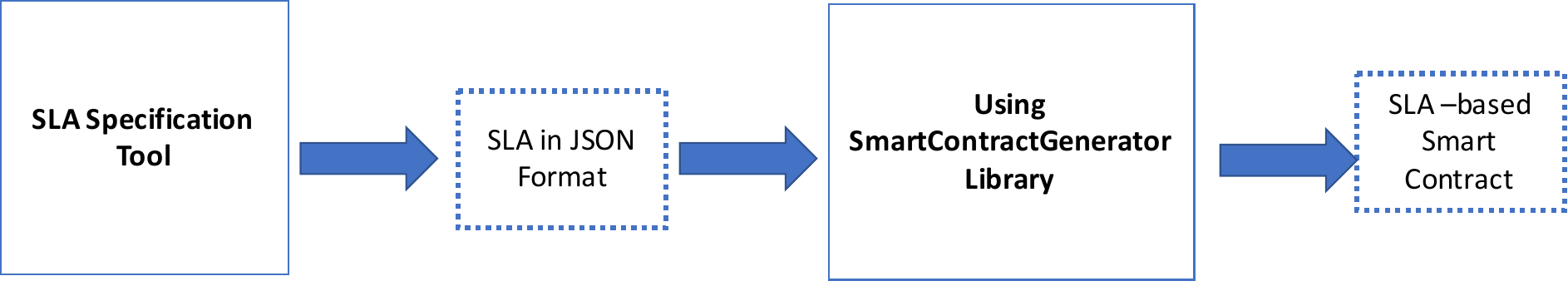}
      \caption{Abstracted generated smart contract from SLA specification steps
}
      \label{fig:FromSLAtoSmart}
   \end{figure*}

A Java library  \footnote{It is explained in Section \ref{sec:lib}} which we refer to as "FromSLAToSmartContract" library. The library converts the generated SLAs for different IoT application use cases with different SLO constraints  to a smart contract (see Figure \ref{fig:FromSLAtoSmart}).\\

 \FloatBarrier
   \FloatBarrier

 Applying FromSLAToSmartContract library to the SLA, can create a list of rules for each SLO constraint and configuration requirement related to each workflow activity.  


\subsection{FromSLAToSmartContract Library} \label{sec:lib}
With the structure of the JSON SLA produced now in mind, the structure of the Java library created can be more easily understood. Since the library targets the SLA generation tools resulting SLA's, the data model of the library is heavily influenced by that of the SLA tool.
\FloatBarrier
\begin{figure}[!htbp]
	\center{\includegraphics[width=\textwidth,height=9cm,keepaspectratio]{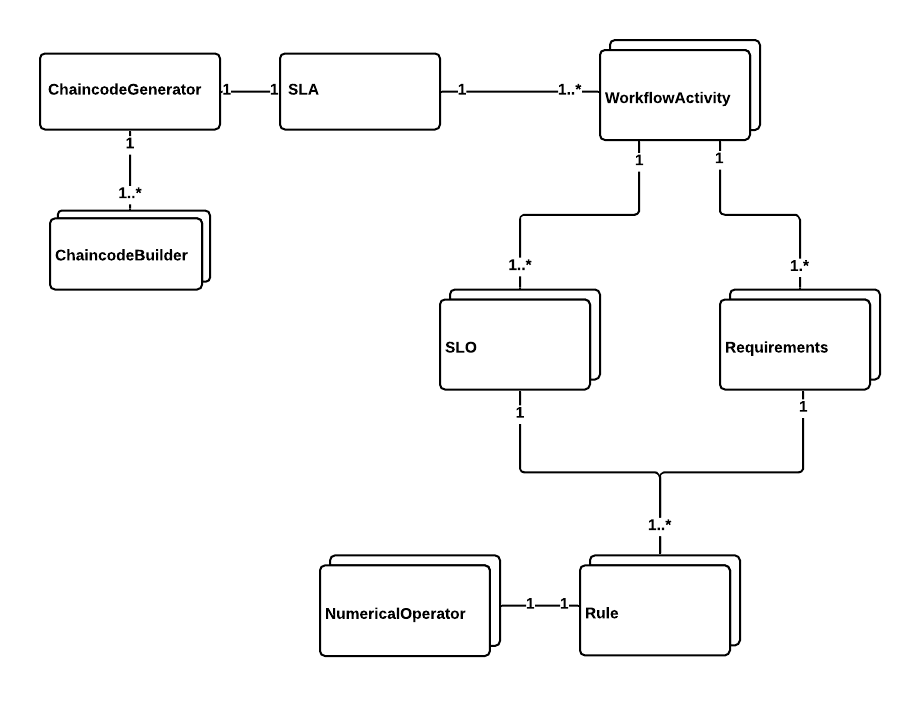}}
	\caption{\label{fig:libstruct} Entity-Relationship diagram of the components in the library}
\end{figure} 
\FloatBarrier

\begin{figure}[!htbp]
	\center{\includegraphics[width=\textwidth]{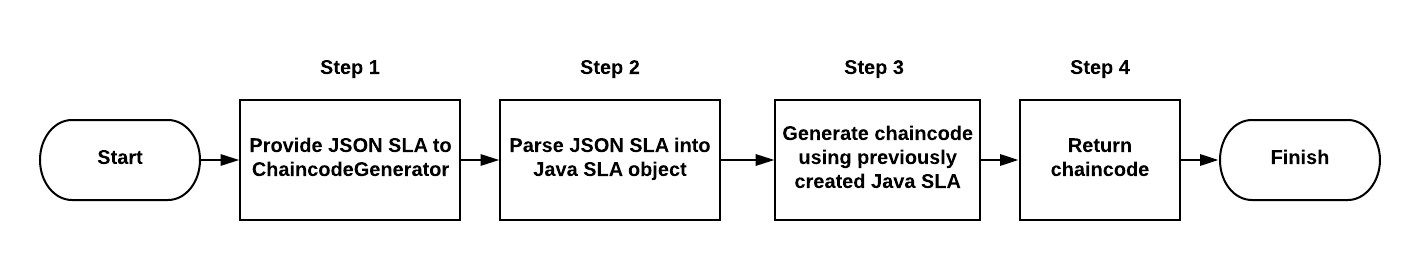}}
	\caption{\label{fig:absslagen} From JSON to Java Chaincode flow}
\end{figure}

Figure \ref{fig:libstruct} shows an abstract view of the components within the Java library along with the composition of these objects. Types in the diagram above such as ChaincodeBuilder and WorkflowActivity have been shown abstractly, however, there is a concrete type for each workflow activity within the SLA generation tool. and are shown in more detail in section \ref{class_section}.
The following is a very brief description of each component within the library:
\begin{itemize}
 \item  ChaincodeGenerator: The entry point of the library. A user of the library calls a static method within this class providing an SLA as either a file handle or JSON string and returns the generated chaincode.
 \item  ChaincodeBuilder: Each workflow activity has an associated Chaincode- Builder. It is the responsibility of the ChaincodeBuilder to convert the rules within a workflow activity into valid chaincode.
 \item SLA: The SLA type represents the service level agreement. It contains fields for the start and end date along with the various workflow activities and application SLOs.
 \item  WorkflowActivity: Represents a specific workflow activity within the SLA. Each workflow activity contains the various SLOs and configuration  requirements for that activity.
 \item  SLO: The service level objective of a specific service (e.g. batch processing, real-time analysis). It contains the  QoS constraints.
 \item  Requirements: Contains the  constraints for various configuration  requirements.
 \item  Rule: Contains a numerical value of the expected QoS constraint along with a NumericalOperator. These two fields are used in conjunction to determine whether the rule is being violated when tested against a submitted value.
 \item  NumericalOperator: Operators (less than, greater than etc.) which when given two values returns a Boolean result. Provides an object-oriented approach to building Rules following the strategy design pattern.
\end{itemize}
As previously mentioned, the ChaincodeGenerator class is the "entry point" to the library. It is this class that a user will interact with by providing a JSON SLA document which then returns a smart contract capable of monitoring the IoT system for SLA violations. Figure \ref{fig:absslagen} provides an abstract view of this process.
\FloatBarrier
\FloatBarrier

The major aspects of the library are steps 2 and 3 i.e. parsing the JSON into a Java object and then generating chaincode based on the object created. With these two tasks being somewhat complex, the following two subsections explain these processes.
\subsubsection{Parsing a JSON SLA into a Java SLA}
\begin{figure}[!htbp]
	\center{\includegraphics[width=\textwidth,height=9cm,keepaspectratio]{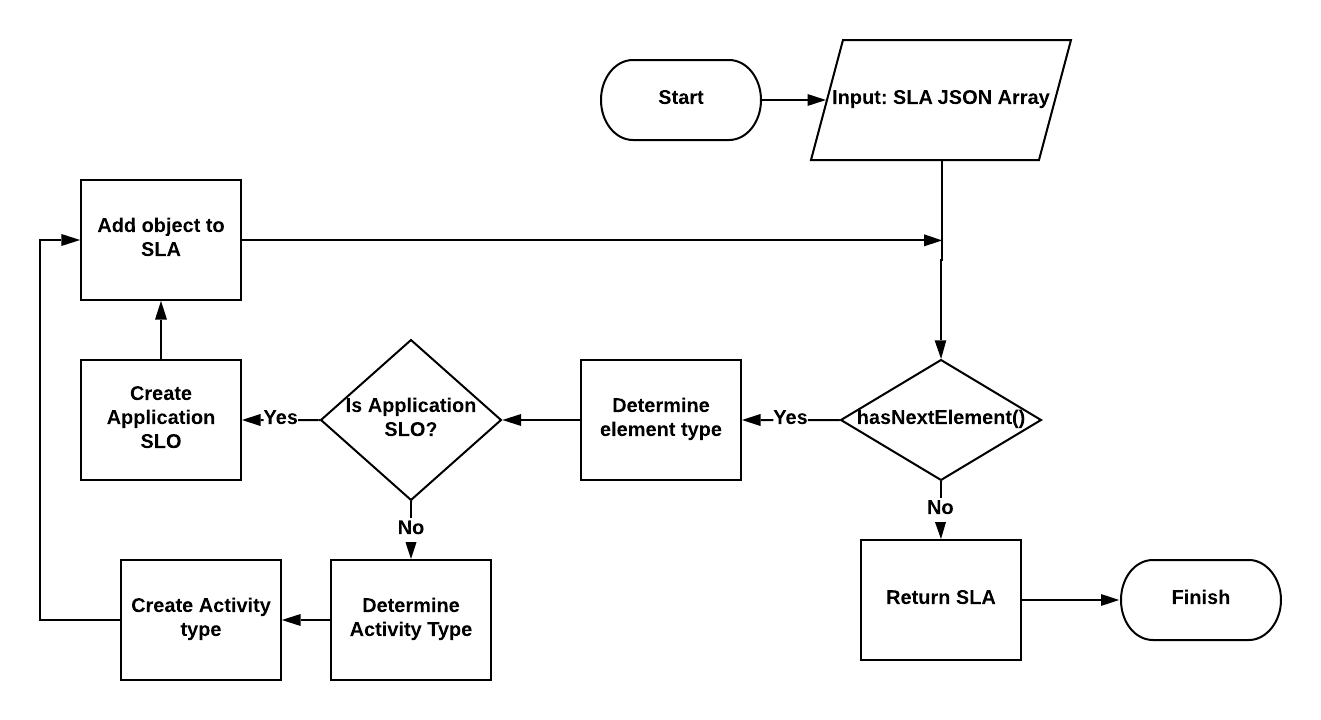}}
	\caption{\label{fig:slacreation} Creating each member of the SLA iteratively}
\end{figure}

The Java SLA type contains a field for each of the possible workflow activities and application SLO defined by the SLA generation tool. This section explains how a JSON document is deserialised into a Java SLA object.
As previously explained, the JSON SLA document produced by the SLA generation tool is a JSON array containing zero or more workflow activities and the application SLO. Therefore, to create a Java SLA object, each of the workflow activity types and application SLO must be determined and constructed. This process is illustrated in figure \ref{fig:slacreation}.
\FloatBarrier
\FloatBarrier

\begin{figure}[!htbp]
	\center{\includegraphics[width=\textwidth,height=9cm,keepaspectratio]{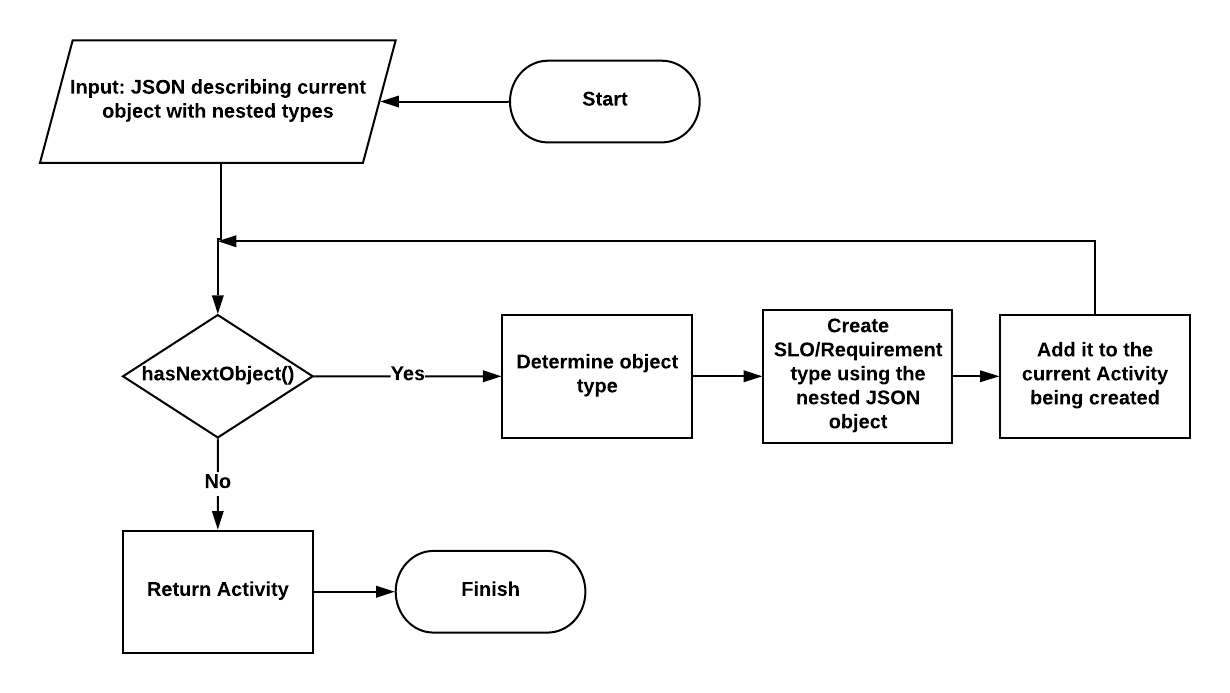}}
	\caption{\label{fig:creatingworkflow} Creating a workflow activity process}
\end{figure}

With each workflow activity and the application SLO being an object within the array, each JSON element of the array is passed to the factory method of the respective workflow activity type. As shown in section \ref{class_section}, each workflow activity in the library contains members for each of the possible SLO’s and Requirements types. When calling the factory method of a workflow activity and providing the JSON object for that activity, the process illustrated in figure \ref{fig:creatingworkflow} is carried out.
\FloatBarrier

\FloatBarrier

\begin{figure}[!htbp]
	\center{\includegraphics[width=\textwidth,height=7cm,keepaspectratio]{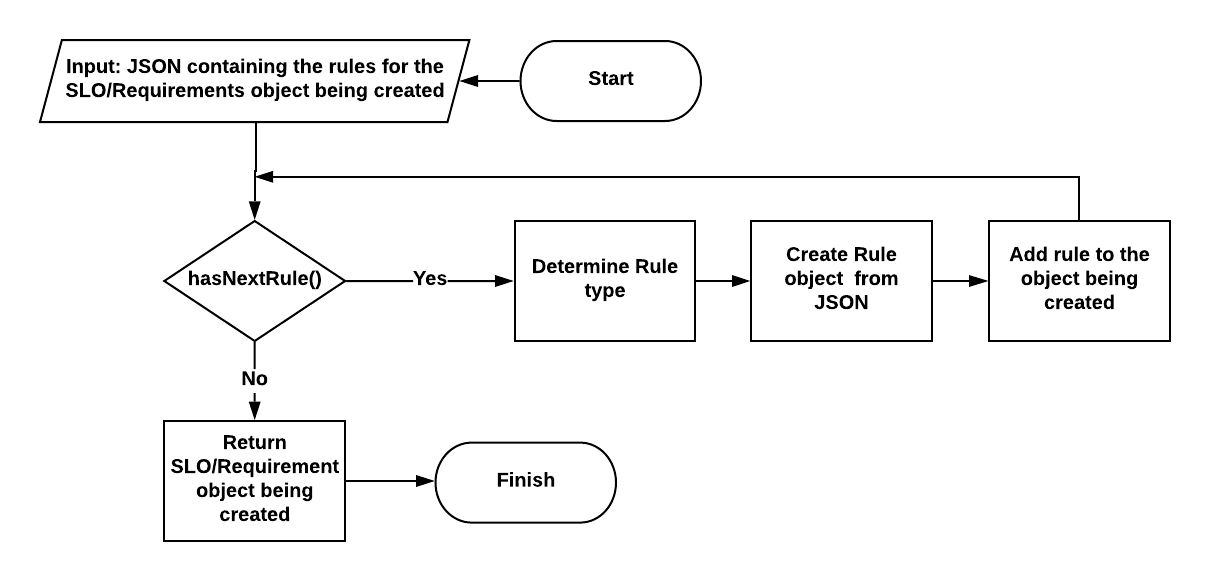}}
	\caption{\label{fig:creatingslo} Creating an SLO/Requirement object}
\end{figure}

The priority of the factory method of the activity type is to determine what SLO’s and Requirement objects are within the JSON object provided, creating these objects in Java and setting them as a member of the workflow activity. An SLO/Requirement object is created by providing the JSON of the type discovered to the relevant SLO/Requirement static factory method which follows the process shown in figure \ref{fig:creatingslo}.
\FloatBarrier

\FloatBarrier

\begin{figure}[!htbp]
	\center{\includegraphics[width=\textwidth,height=9cm,keepaspectratio]{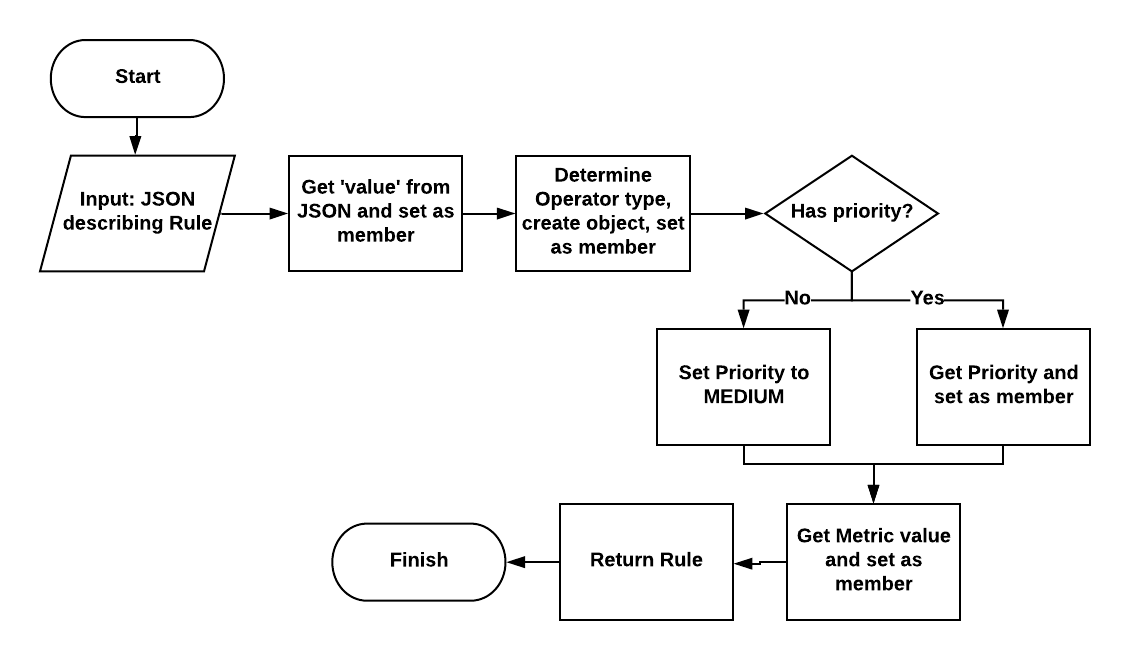}}
	\caption{\label{fig:createrule} Create Rule flow}
\end{figure}

The static factory method of the SLO/Requirement type inspects the JSON provided for any rules relevant to the type being created and creates this Rule in Java by providing the JSON to the static factory method of the appropriate Java Rule class. A Rule is created by following the process shown in figure \ref{fig:createrule}.
\FloatBarrier

\FloatBarrier

As shown in the previous flow diagrams, creating a Java SLA is somewhat of a recursive and iterative process. Each workflow activity within the SLA must be created and is achieved by the calling factory method for that activity type and providing the JSON describing it. The workflow activity is constructed by creating all SLO/Requirement objects described in the JSON passed to the activity factory method. This is achieved by providing each nested JSON object describing the SLO/Requirements to the appropriate factory methods of each type. The factory method of the SLO/Requirement creates each Rule described in the JSON passed to its factory method.
Once a Java SLA object has been created, the library progresses to the next step which generates chaincode based on the various objects defined in the SLA.
\subsubsection{Generating Chaincode from Java SLA}
As shown in the previous subsection, an SLA object within the smart contract generation library consists of zero or more workflow activities and an application SLO each containing Rules. It is these "Rules" that express the required QoS metrics for the IoT system and are used to validate the state of a system being monitored.

Each workflow activity and application SLO type have an associated ChaincodeBuilder class. A ChaincodeBuilder extracts each SLO/Requirement object from the provided workflow activity and then generates three chaincode methods for each of the SLO/Requirement types.
The three methods generated for each SLO/Requirement objects are:
\begin{itemize}

\item*\_update: The update method used to report the current state of the SLO/Requirement being captured in the IoT system. E.g. "application\_slo\_update".
\item get\_latest\_*\_update: Returns the latest state reported to the smart contract. E.g. "get\_latest application\_slo\_update".
\item get\_*\_violations: Returns all recorded violations of the SLA for that specific SLO/Requirement.\\ E.g. "get\_application\_slo\_violations".
\end{itemize}
Each ChaincodeBuilder generates these three methods for each SLO/Requirement object by following the process illustrated in figure \ref{fig:genchaincode} when the getMethods() method is called:
\FloatBarrier

\begin{figure}[!htbp]
	\center{\includegraphics[width=\textwidth,height=9cm,keepaspectratio]{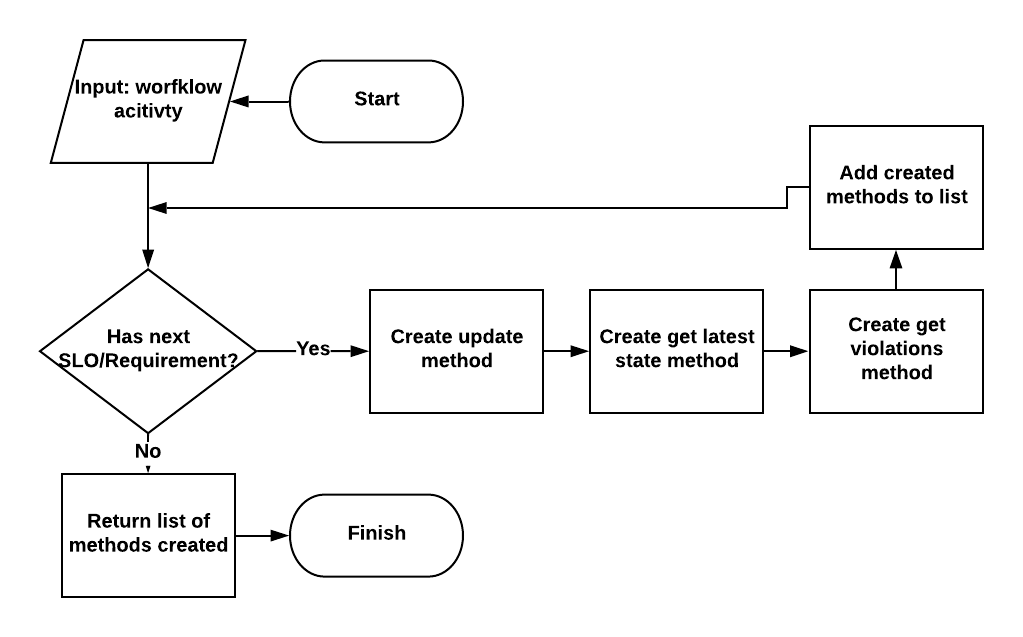}}
	\caption{\label{fig:genchaincode} Constructing chaincode methods using a ChaincodeBuilder}
\end{figure} 
\FloatBarrier

The update methods are created by extracting each of the QoS rules within the SLO/Requirement objects and creating if statements to check against submitted values. For example, if an SLO contains a rule for availability stating that it must be always greater than 99 per cent, the if statement contained within the update method would read:\\
$if ( !(availability > 99)) {\\
***record_violation***\\
}$\\
This is achieved with the assistance of the Java library "JavaPoet" \cite{28}  to construct a method and add control statements using the Rule object values. For each rule in an SLO/Requirement object, a corresponding if statement is generated and contained within the same update method.
The ChaincodeGenerator class is the entry point for the library and acts as the API. The ChaincodeGenerator class takes advantage of the ChaincodeBuilders described above by using these objects to generate a list of methods for all workflow activities within the Java SLA object. Once all methods have been generated, the ChaincodeGenerator class simply creates a Java class using the functionality provided by JavaPoet and appends these methods to the class. Finally, the ChaincodeGenerator returns a Chaincode object containing the Java class as a string and also documentation for the generated contract. A fully deployable Hyperledger Fabric smart contract project is also written to the file system, allowing the user to directly deploy the smart contract to the blockchain.
\subsubsection{Chaincode}
The smart contract generated by the library contains three methods for each SLO/Requirement objects and also three other methods which are "init", "invoke" and "history".
The init method is called by Hyperledger Fabric when the smart contract is initialised and should contain any initialisation logic required for the contract. However, in the case of the monitoring contracts that are generated, there is no initialisation logic required and therefore this method is empty.\\

The invoke method is the external API of the smart contract and is the only method the end user will directly interact with. The invoke method expects two string parameters with the first being the name of the internal method to invoke e.g "batch processing slo update" and the JSON string containing the state of the batch processing SLO.
The invoke method simply contains a switch statement containing a case for each method in the chaincode. When invoke is called, the method name submitted in the first parameter is evaluated by the switch statement and calls the relevant method providing the JSON string to it. If the method name is not recognised, an error is returned.\\

When an update method is called, the JSON string provided to it is converted into a Java JSON object, allowing for the values contained within the JSON to be extracted. Each update method expects the JSON to contain a key/value pair for each rule defined in the SLA along with an ID of the resource that produced it. The update method compares each value found within the JSON against the rules defined in the SLA for that type. If a value is missing, an error is recorded and is viewed as a violation for not being included.\\

If a value submitted to the update method violates a QoS rule, it is recorded as a violation on the blockchain. The violation record contains the rule that was violated, the value that caused the violation, the ID of the resource that produced the value and a timestamp. The record is stored under a composite key of both the SLO/Resource type and the ID of the resource that produced the state. For example, the violations recorded for a Gateway SLO produced by a device with ID 1 is a separate set from the violations produced by a Gateway SLO with the ID of 2.\\

After evaluating the submitted state for violations, the state is then stored in the ledger and can be read back using the get latst\_*\_update method. Similarly, a list of all violations for a specific SLO/Requirement associated with an object ID can be retrieved using the get\_*\_violations method. Listing \ref{fig:invoke} shows a simple example of the invoke method generated:
\FloatBarrier
\begin{lstlisting}[language=json,firstnumber=1, basicstyle=\linespread{0.8}\fontsize{10}{10}\selectfont\ttfamily, caption={ "Generated invoke method"},label={fig:invoke}]
public org.hyperledger.fabric.shim.Chaincode.Response invoke(ChaincodeStub stub) {
    String method = stub.getFunction();
    List<String> params = stub.getParameters();
if (method.equalsIgnoreCase("examine_captured_eoi_gateway_slo_update")) {
      return examine_captured_eoi_gateway_slo_update(stub, params.get(0));
    }
    if (method.equalsIgnoreCase("get_latest_examine_captured_eoi_gateway_slo_update")) {
      return get_latest_examine_captured_eoi_gateway_slo_update(stub, params.get(0));
    }
    if (method.equalsIgnoreCase("get_examine_captured_eoi_gateway_slo_violations")) {
      return get_examine_captured_eoi_gateway_slo_violations(stub, params.get(0));
    }
return newErrorResponse("ERROR", "Invalid function name".getBytes());
  }



\end{lstlisting}
\FloatBarrier

In this example, only the code is shown the for gateway SLO belonging to the examine captured EoI activity. When an update for this SLO is passed to the invoke method is executed.

In this example, only the code is shown the for gateway SLO belonging to the examine captured EoI activity. When an update for this SLO is passed to the invoke method, the code illustrated in Listing \ref{fig:27}  is executed.
\FloatBarrier
\clearpage
\begin{lstlisting}[language=json,firstnumber=1,basicstyle=\linespread{0.8}\fontsize{10}{10}\selectfont\ttfamily,
 caption={ "Generated smart contract update method"},label={fig:27}]
private org.hyperledger.fabric.shim.Chaincode.Response application_slo_update(ChaincodeStub stub,
      String jsonString) {
    try{
    JsonObject json = new JsonParser().parse(jsonString).getAsJsonObject();
    byte[] state = stub.getState("APPLICATION_SLO_VIOLATION");
    JsonArray arr;
    if (state == null || state.length == 0) {
      arr = new JsonArray();
    }
    else {
      arr = new JsonParser().parse(new String(state)).getAsJsonArray();
    }
if (json.has("GATEWAY_AVAILABILITY")) {
      double val = json.get("GATEWAY_AVAILABILITY").getAsDouble();
      if (!(val >= 99.9)) {
        String msg = "GATEWAY_AVAILABILITY rule violated. Offending value is: " + val + " at time " + new Date().toString();
        arr.add(msg);
      }
    }
    else {
      String error = "GATEWAY_AVAILABILITY" + " missing at time " + new Date().toString();
      return newErrorResponse("ERROR", error.getBytes());
    }
stub.putState("EXAMINE_CAPTURED_EOI_GATEWAY_SLO_VIOLATIONS" + "_" + id, arr.toString().getBytes());
    stub.putState("EXAMINE_CAPTURED_EOI_GATEWAY_SLO" + "_" + id, jsonString.getBytes());
    return newSuccessResponse("EXAMINE_CAPTURED_EOI_GATEWAY_SLO_UPDATE", jsonString.getBytes());
    } catch (Exception e) {
    return newErrorResponse("ERROR", e.getMessage().getBytes());
    }
  }



\end{lstlisting}
\FloatBarrier

The Gateway SLO has one rule in this example which states that the availability should always be above 99.9 per cent. The method exacts the gateway availability value from the JSON string submitted and compares it against the expected value. If the availability is recorded below 99.9 per cent, a violation is recorded on the ledger. The update is added to the ledger regardless of violations and is returned to the caller.\\

Lastly, the entire history of updates for a specific SLO/Requirement can be read back by using the invoke method and providing the string "history" and the name and ID of the SLO/Requirement to get the history for. When this occurs, a list of all historic updates are returned for that aspect of the system with each update containing the timestamp of when the update occurred.

\section{Evaluation}
In this section, a fictitious Remote Patient Monitoring (RPM) IoT system is presented along with a service level agreement defined for the system. A smart contract is generated using the smart contract generation library developed in this project which is then deployed to a Hyperledger Fabric network forming a consortium blockchain. The smart contract is then tested using an IoT emulation program developed to simulate various structures of the RPM IoT system and the results of these tests are presented.
\subsection{Scenario}
The example scenario presented in this section is in the form of a Remote Patient Monitoring (RPM) IoT system that monitors the heart rate and blood pressure of patients due to some cardiovascular disease. The system is used to detect early signs of serious heart issues/defects and deploys emergency services in the case of heart failure. While the IoT system presented is fictitious with its purpose to demonstrate the usage of the smart contract generation library, it does however take inspiration from existing research in remote health monitoring such as \cite{26}. \\

\begin{figure}[h!]
	\center{\includegraphics[width=\textwidth]{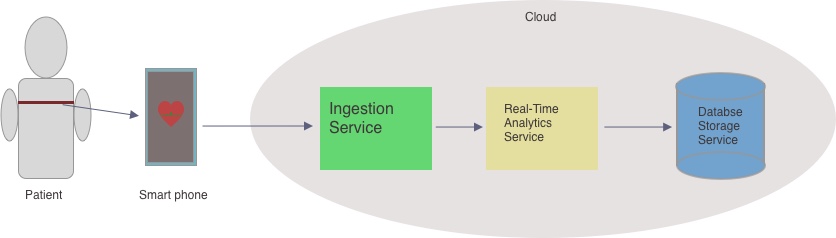}}
	\caption{\label{fig:trpm} RPM IoT System}
\end{figure}

Figure \ref{fig:trpm} illustrates the RPM IoT system responsible for monitoring and detecting heart failure and other serious cardiovascular health issues. The first layer (Patient) shown in the diagram is the sensor layer, which is responsible for reading the heart rate and blood pressure every 10 seconds of the patient. The edge device layer (smart phone) is a smart phone that collects heart rate and blood pressure readings from the sensors, filters this data and aggregates it to the ingestion layer of the IoT system. The ingestion layer is responsible for receiving data from many patients, which is then processed and provided to the real-time data analytics service. The real-time data analytics service analyses the data provided looking for health issues and heart failure for each patient. If it detects some issue, it notifies health care professionals which may result in emergency services being deployed to those in need. Finally, from the real-time data analytics service the data is stored in a database that can later be queried by medical staff.
The IoT system consists of three organisational members including two service providers (SP1 and SP2) that are providing hardware infrastructure (e.g. sensors) along with services such as cloud-based data storage. The final organisational member is the health care provider (HCP) who uses the system to provide monitoring and health support for their patients.
\FloatBarrier

\FloatBarrier

\subsection{Service Level Agreement}
With the IoT system being a type of smart-health application and is responsible for detecting serious health issues in real time, it is crucial that this system works correctly and to the desired level of service. If the system fails to meet the demanded quality of service it may result in serious consequences such as deaths that could have been prevented along with lawsuits.\\

An SLA was created using the SLA generation tool \cite{10}  describing the required service level objectives and resource requirements for each layer of the RPM IoT system. The service level agreement includes required QoS rules of the services used such as the required availability, accuracy and cost. Each organisation that provides functionality (SP1, SP2) agree to the SLA stating that the services they provide meets this standard and is therefore safe to use. Failure to meet the required service level agreement will incur legal action against those who fail to provide services at the agreed upon standard. With all organisations agreeing to the defined SLA, the IoT system is deployed. 
\subsection{Smart Contract Generation}
With an SLA defined, the next task is to generate a smart contract capable of monitoring the deployed system for SLA violations. Generating the smart contract is achieved by creating a simple Java program that references the smart contract generation library created in this project. This program simply calls the ‘chaincodeFromFile’ method within the smart contract generation library and provides a file reference to the JSON SLA produced by the SLA generation tool (previous step) and a path for where the generated smart contract project should be written to. After executing this program, the following files and folders were created resulting in a deployable smart contract project:\\
\begin{figure}[h!]
	\center{\includegraphics[width=\textwidth,height=4cm,keepaspectratio]{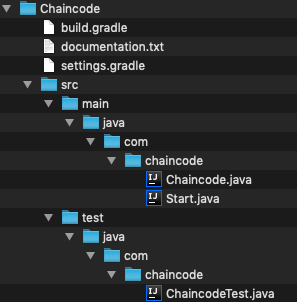}}
	\caption{\label{fig:rpm} Generated Smart Contract Project}
\end{figure}

The chaincode folder contains all files required to form a deployable smart contract to a Hyperledger Fabric network. The Start.java class is the entry point for the smart contract that the Hyperledger Fabric platform will execute to start the chaincode application that then loads the Chaincode.java class. The Chaincode.java class is where the generated smart contract resides. This class contains all generated code which an end user/application will interact with to update various states and to read any violations. It is these update methods that will evaluate the submitted states to check whether the SLA is being violated and record these states and violations to the distributed ledger.\\

The ChaincodeTest.java file contains unit tests for the generated chaincode, however, these unit tests do not include any test logic for the smart contract and are instead provided if a user wishes to write their own tests. The ‘build’ and ‘settings’ gradle files are used by Hyperledger Fabric peers to build and deploy the chaincode project.\\

The chaincode generated from the RPM IoT SLA contained over 2,300 lines of code consisting of 57 methods. The chaincode was manually inspected and the validation logic generated was correct according to the defined SLA.
\subsection{Chaincode Deployment}
\begin{figure}[h!]
	\center{\includegraphics[width=\textwidth]{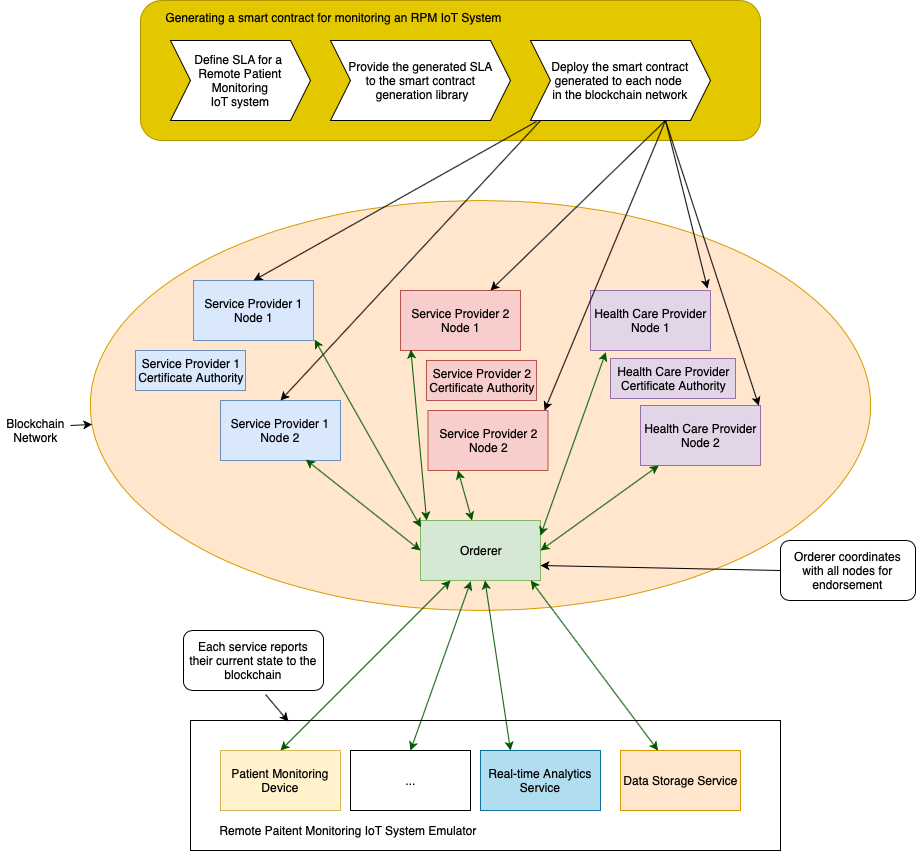}}
	\caption{\label{fig:testnetwork} Test Network}
\end{figure} 
With a Hyperledger Fabric chaincode project generated, the next task was to deploy a Hyperledger Fabric network that would be used to monitor the deployed RPM IoT system.
The network used in this example takes advantage of the "first-network" \cite{27}  sample provided by Hyperledger Fabric to deploy a consortium blockchain consisting of three organisations (SP1, SP2 and HCP). The purpose of this network was to demonstrate the functionality of the smart contract generated and does not intend to accurately benchmark Hyperledger Fabric blockchain technology.\\

Therefore, this network is being used for testing the smart contract generated and all peers are deployed locally. In a production context, however, a network would be deployed to many machines with each organisation running their own peers on their own hardware.
While this text briefly explains the deployment of the network used in this example along with installing and instantiating chaincode, the reader is encouraged to read \cite{27}. 
The manual creation of Hyperledger Fabric networks is somewhat laborious and complex and is out of scope for this text.\\
 
Hyperledger Fabric provides a project called "fabric samples" which includes many tools and networks such that a user can easily deploy a prototype network on their local machine. One of these sample networks provided is called "first- network" which a user can execute two scripts and deploy a Hyperledger Fabric consortium network consisting of three organisations. This process was carried out and after executing the "byfn" and "eyfn" scripts within the first-network project and providing a channel name of "healthiot" along with the "-a" parameter as arguments to these scripts, the network was deployed.\\

Once the network was deployed, the smart contract was installed and instantiated on each of the peers for each organisation. This was achieved by first copying the generated smart contract project into the chaincode folder of the "fabric-samples" project and then connecting to the Fabric CLI container that was deployed during the deployment of the network. After connecting to the fabric CLI container,  appropriate commands were executed to install and instantiate the smart contract to each peer of each organisation in the network.\\

Figure Figure \ref{fig:testnetwork}  depicts the final network deployed and process taken. The orderer node is responsible for ordering the transactions such that all peers in the network have executed the same transactions in the same order on their local ledger. While there are various distributed orderer protocols, the network deployed uses the solo ordering protocol. Also used is the "goleveldb" database technology to store the recorded state of the ledger.
\subsection{IoT Emulation}
With a Hyperledger Fabric network deployed and the smart contract generated running on each peer, the next step was to test the chaincode and network. Rather than actually deploying a real RPM IoT system, instead an emulator was developed such that it would submit transactions to the blockchain network for each layer of the proposed topology.\\

The emulator was developed using NodeJS and the JavaScript programming language. The emulator takes advantage of JavaScript's asynchronous programming model, allowing multiple transaction requests to be sent without blocking for a response. This therefore allows the emulator to submit transactions to the blockchain network in fast succession in an attempt to simulate real-world transaction speeds.\\

The emulator contains fields configurable by the user to set the number of sensors, gateway devices, real-time analysis services etc. allowing the user to simulate different structures of the IoT system with varying numbers of each service. The emulator would submit state update transactions for each element of the defined IoT system. For example, if the user configures the emulator to emulate an IoT system with 100 sensors at the sensor layer, 100 sensor state update transactions will be submitted to the Fabric network per iteration, thus emulating the deployed system.\\

Not only does the emulator include fields that allows the user to emulate different infrastructures but also contains another field ‘VIOLATION RATIO’. This is a double value ranging from 0 to 1 that allows the user to control how many elements for all service layers will cause a violation. For example, if a user has configured the emulator as 100 sensors at the sensor layer and a violation ratio of 0.05, 5 sensor states containing an SLA violation will be submitted per iteration.\\

The emulation logic is contained within a loop such that a user can configure the emulator to run X iterations. For each iteration, state transactions will be submitted for each defined element within the configured IoT topology in the emulator. For example, if the user has configured the emulator to contain 100 sensors, 50 gateway devices, 5 ingestion services, 3 real-time analytics services and 2 storage services, SLO/Requirement state update transactions will be executed for all 160 elements of the IoT system per iteration. The total number of transactions submitted will depend on the number of SLO/Requirement updates for each device/service type. For example, at the sensor layer there is only one update method, however, the gateway layer requires 4 update methods.
\subsection{Results}
Using the emulator described previously, the network and smart contract deployed was tested using various emulated structures of the proposed RPM IoT system (see Table \ref{tab:result}). Each test was performed using a violation ratio of 0.05 with at least one element of each layer producing a violation. Each test ran for three iterations, with the average time being recorded as the performance time. The tests were executed on a PC equipped with an i7-5820K CPU with 16GB RAM running Debian 9 with the Hyperledger Fabric network deployed locally.\\

The purpose of these tests were to check that the deployed smart contract would detect all SLA violations produced by the emulator and that each submitted state was recorded to the ledger. While the average time is shown in the results, this is not an accurate representation of Hyperledger Fabrics transaction processing ability. It is also worth noting that no efforts were made to optimise the deployed network.

\begin{table}[]\label{tab:result}
\begin{tabular}{|l|l|l|l|l|l|l|l|}
\hline
\multicolumn{8}{|c|}{Performance of the Generated Smart Contract with RPM IoT Emulation}                                                                                                                  \\ \hline
\#Sensors & \#Gateway & \#Ingest & \#RT Analytics & \#Storage & Avg. Time & \begin{tabular}[c]{@{}l@{}}\#Trans.\\  Performed\end{tabular} & \begin{tabular}[c]{@{}l@{}}Violations\\ Detected\end{tabular} \\ \hline
2         & 1         & 1        & 1              & 1         & 2.5s      & 18                                                            & 17/17                                                         \\ \hline
20        & 10        & 1        & 1              & 1         & 3.32s     & 72                                                            & 21/21                                                         \\ \hline
100       & 50        & 1        & 1              & 1         & 6.95s     & 317                                                           & 29/29                                                         \\ \hline
200       & 100       & 2        & 2              & 1         & 20.23s    & 620                                                           & 42/42                                                         \\ \hline
500       & 250       & 3        & 3              & 2         & 39.58s    & 1,532                                                         & 89/89                                                         \\ \hline
1,000     & 500       & 5        & 5              & 4         & 71.54s    & 3,056                                                         & 162/162                                                       \\ \hline
\end{tabular}
\end{table}
\section{Discussion}
The example scenario and results presented in the previous section represent a proof-of-concept by demonstrating the functionality of the auto-generated smart contract derived from an SLA. While the tests do show that the generated smart contract does in fact capture all violations and records them on the distributed ledger, it fails to demonstrate Hyperledger Fabric as being an optimal choice for IoT monitoring. However, this is not to say that Hyperledger Fabric is not suitable.\\

The tests shown previously illustrate that Hyperledger Fabric was slow in processing the transactions, especially as the emulated system was scaled. However, the Fabric network was deployed locally with all peers running on the same machine and did not fairly represent a production standard Fabric network. Therefore, the speed (transactions per second) in this case should not be seen as a reliable metric of the system.\\

With the project utilising the network sample provided by Hyperledger Fabric, there was no optimisation performed on the network. Research such as \cite{29} \cite{30}  suggest that various approaches can be taken to optimise a Fabric network, implying that it may be possible to achieve much higher transactions per second, possibly making Hyperledger Fabric an appropriate blockchain for IoT monitoring. Therefore, further research is required into optimising and benchmarking Hyperledger Fabric running the smart contracts generated by the library to determine the suitability of Hyperledger Fabric in a production setting.\\

The IoT emulator did function correctly and submitted the correct number of transactions for each element of the IoT system along with the correct number of violations. However, the emulator submitted these transactions sequentially per iteration as opposed to concurrently. While the emulator takes advantage of asynchronous programming, meaning that once the transaction is submitted it is capable of submitting another without blocking, nevertheless, these transactions are not submitted concurrently and do not accurately emulate an IoT system. This is due to IoT systems consisting of many individual components each submitting transactions possibly at the same time, whereas the emulator is unable to do this. Despite not emulating the concurrent nature of an IoT system, however, the emulator does submit these transactions in very fast succession and does demonstrate the smart contracts ability to detect violations.\\

After each test performed on the example network and smart contract, the distributed ledger was examined to verify that all SLA violations were captured and recorded to the ledger along with the state updates for each SLO/Requirement type. During each test the smart contract did capture all violations and recorded them along with recording all states submitted, thus demonstrating the correctness of the smart contract generated.\\

Consortium blockchains provide distributed ledger technology such that all members of the consortium have the same data. It is this distributed property along with the immutability of the ledger that makes it an attractive option when monitoring an IoT system for SLA violations with multiple parties. During the process of developing the smart contract library however, some potential flaws in using blockchain technologies for this purpose were discovered.\\

The goal of the smart contract generated is to detect SLA violations which in turn may cause a member of the consortium to be penalised for not meeting the requirements of the SLA. Therefore, it is within a members interest for a violation they are committing to not be recorded on the distributed ledger. In some cases, it may be a requirement of a blockchain network that all members of the consortium accept a transaction for it to be recorded on the ledger. If this requirement is in place, it may be possible for the offending member to reject the transaction resulting in the violation not being recorded. While this rejection may be detectable by other members of the consortium with other means, nevertheless, the violation will not be recorded on the expected ledger. In order to solve this issue, the number of accepting members to allow a transaction to be committed to the ledger may be reduced, however, this may then lead to the next possible issue.\\

If the consortium requires less than the number of members within a consor- tium to validate and accept a transaction, it would then be possible for a group of members within the consortium to "team up" on another member and record transactions suggesting that the victimised member did not meet the SLA. This, in turn, would defeat the purpose of using blockchain technology in this setting as it would allow a subset to gang up on a member and generate fictitious SLA violations and therefore requires some level of trust among peers.
Another potential issue of using blockchain technology to monitor IoT systems relates to how the blockchain acquires the system state information of the IoT system. In this project it was assumed that all services/devices would push their current state to the smart contract, thus placing implicit trust on these services/devices. This trust is required due to the possibility of the device reporting false/inaccurate state information, possibly programmed by a malicious member of a consortium wishing to hide their services true operational state. Therefore, a solution to the issue of determining the state of an IoT service in a trustworthy manner is required.\\

To the authors knowledge, there does not exist an API/standard definition for IoT metrics such that all IoT services/devices can publish their state in a standard way. For example, within the smart contract generation library, it is assumed a sensor or "thing" will publish its availability under the JSON key AVAILABILITY. However, this places a strict requirement on the IoT device such that the availability must be reported under this key within a specific JSON format, resulting in a very specific API. A better approach would be to define an industry standard such that all services/devices will emit their state (if monitoring is enabled) in a standard way such that the smart contract generated by the library will be compatible with any service/device within any IoT system. This, in turn, would also solve the tight coupling issue between the smart contract generation library and the SLA generation tool.
\subsection{Limitation}
Although the smart contract generation library developed does generate valid smart contracts that accurately detects SLA violations, the structure of the library could be improved. The library developed is tightly coupled to the SLA generation tool along with the JSON SLAs that the tool produces. This can be attributed to the data model and SLA deserialization code.\\

The data model of the library is heavily influenced by that of the SLA generation tool. This was due to modelling the various domain objects such as workflow activities and SLO/Requirement objects based on the structure and information found within the graphical user interface (GUI) of the SLA generation tool. Therefore, the smart contract generation library expects certain workflow activities to have specific SLO/Requirements types and if, for example, the SLA generation tool was to be updated to include a new rule for the gateway SLO, these changes would need to be reflected in the smart contract generation library. This tight dependency on the SLA generation tool causes the smart contract generation library to be brittle, meaning that any changes to the SLA tool may cause the library to fail. Removing the tight coupling between the SLA generation tool and smart contract generation library would improve the library and make it less brittle. Achieving this may require further research into defining service level agreement standards such as SLA keys so that it is safe to assume they will rarely (if ever) change along with defining an industry standard for a JSON structured SLA.
\section{Conclusion and Future Work}
Monitoring IoT systems for SLA violations presents interesting challenges, especially when multiple organisations are involved and trust is difficult to achieve. This research project aimed to explore the possibility of using blockchain technology and smart contracts to monitor an IoT system in order to achieve implicit trust among all parties that are monitoring the system.\\

A Java library was developed that could convert an end-to-end IoT SLA into a smart contract that can be directly installed onto a Hyperledger Fabric consortium blockchain. The smart contracts generated by the library can detect any SLA violation and record the violation to the distributed ledger, thus making the data captured immutable. The smart contracts also record each reported state to the ledger such that a member of the consortium can view all updates recorded and can query the blockchain for all violations of a specific aspect of the IoT system.\\

An SLA was created to represent a remote patient monitoring system and the library developed was then used to generate a smart contract from this SLA. A Hyperledger Fabric blockchain network was deployed locally, consisting of three organisations that represented the consortium of the IoT system. Once the generated smart contract was installed on to the blockchain, a simple IoT emulator was developed using NodeJS that would submit state transactions to the blockchain with some containing SLA violations to test the smart contract functionality.\\

While testing the smart contract on a local blockchain network, multiple IoT system structures were emulated, testing how well the smart contract could handle a greater number of nodes in a system. It was noted that as the emulated system scaled up, the time to capture the state also increased with the final test taking over 70 seconds per full iteration. While this speed may be an issue for large-scale IoT systems, it is worth noting that the smart contract did detect and record all SLA violations that were generated during these tests.\\

Although the smart contract generated achieved the functional goal of detecting all SLA violations, further research was identified. Firstly, the transaction throughput of the blockchain may not be fast enough for large IoT systems and therefore exploring ways to increase this speed or to reduce the number of transactions for monitoring is required. Research such as \cite{29} and \cite{30}  have suggested that it could be possible to achieve a higher transaction throughput using Hyperledger Fabric with various parameters and optimisation techniques, however, this was not carried out in this research.\\

The Java library developed is also heavily dependent on the end-to-end IoT SLA generation tool such that it relies on the JSON output of the SLA generation tool. Removing this coupling issue is required and could be solved by defining an industry standard of keys/names for specific aspects of an IoT system such that all future IoT systems and SLA generation tools would report their state under the same keys.\\

The smart contract generation library only targets Hyperledger Fabric, pro- ducing smart contracts in the Java programming language. Due to the design of the library, support for other blockchain platforms and smart contract languages could be added, thus catering for a wider audience.
The smart contracts generated by the library developed have been demonstrated to capture all SLA violations and could be a step closer to achieving an IoT monitoring solution to increase trust and remove the possibility of foul play among organisations forming a consortium network.\\

As future work, Currently the library created only generates smart contracts that detects SLA violations and records them to the ledger. However, having the smart contract automatically issue a penalty when a violation occurs would further increase the ability of the smart contract. With the current smart contracts generated by the library, human interaction is still required, however, it could be possible to have the library to generate smart contracts that also mediates SLA violation penalties.\\

The design of the library developed relies strictly upon JSON keys produced by the end-to-end IoT SLA generation tool along with defining its own API for IoT services to submit the current state of the service. Defining a standard that details specific JSON keys for specific values along with an industry standard API for reporting monitoring state would remove reliance upon specific tools. By creating a standard that all future services and tools developed would adopt would therefore remove any coupling issues.\\

Finally, the library created in this project currently only supports a single blockchain platform (Hyperledger Fabric) and only produces a smart contract in the Java programming language. Adapting the Java library to support other blockchain platforms such as Hyperledger Sawtooth and Ethereum may be beneficial to those who have strict requirements to use one of these alternative blockchain platforms.
\bibliographystyle{unsrt}

\bibliography{sample}

\end{document}